\RequirePackage{fix-cm}
\documentclass[twocolumn]{svjour3}          
\usepackage[switch]{lineno} 
\smartqed  
\usepackage{graphicx}
\usepackage{xcolor}
\usepackage{epstopdf}
\usepackage[labelformat=parens,labelsep=quad,skip=3pt,justification=justified, format=plain]{caption}
\usepackage[utf8]{inputenc}
\usepackage[english]{babel}
\usepackage{natbib}
\usepackage{placeins} 
\usepackage{hyperref}
\begin{document}
\sloppy
\title{Extreme events in a polar warming scenario--a laboratory perspective
\thanks{This work was supported by the \emph{Spontaneous Imbalance Project} (HA 2932/8-1 and HA 2932/8-2) that is part of the research group \emph{Multiscale Dynamics of Gravity Waves} funded by DFG (FOR1898). }
}

\titlerunning{Extreme events in a polar warming scenario}        

\author{Costanza Rodda        \and
             Uwe Harlander  	    \and
            Miklos Vincze
}


\institute{C. Rodda \at
Brandenburg University of Technology Cottbus–Senftenberg, Department of Aerodynamics and Fluid Mechanics, Cottbus, D-03046, Germany;\\
              Laboratoire des Ecoulements G\'eophysiques et Industriels, Universit\'e Grenoble Alpes, F-38000 Grenoble, France \\
              \email{costanza.rodda@univ-grenoble-alpes.fr}           
           \and
           U. Harlander \at
              Brandenburg University of Technology Cottbus–Senftenberg, Department of Aerodynamics and Fluid Mechanics, Cottbus, D-03046, Germany
	\and
           M. Vincze \at
           von Ka\'rm\'an Laboratory of Environmental Flows; E\"{o}tv\"{o}s Lor\'and University, Budapest H-1117, Hungary/
           MTA-ELTE Theoretical Physics Research Group; E\"{o}tv\"{o}s Lor\'and University, Budapest H-1117, Hungary
}

\date{Received: date / Accepted: date}

\maketitle

\begin{abstract}
We report on a set of laboratory experiments to investigate the effect of Arctic warming on the amplitude and drift speed of the mid-latitude jet stream. Our results show that a progressive decrease of the meridional temperature difference 1) slows down the eastward propagation of the jet stream, 2) complexifies its structure, and 3) increases the frequency of extreme events. 
Extreme events and temperature variability show a clear trend in relation to the Arctic warming only at latitudes influenced by the jet stream, whilst such trend reverses in the equatorial region south of the subtropical jet.  
Despite missing land-sea contrast in the laboratory model, we find similar trends of temperature variability and extreme events in the experimental data and the National Centers for Environmental Prediction (NCEP) reanalysis data. Moreover, our data qualitatively confirm the decrease in temperature variability due to the meridional temperature gradient weakening (which has been proposed recently based on proxy data). Probability distributions are weakly sensitive to changes in the temperature gradient, which can be explained by recent findings using quasigeostrophic models. 
\end{abstract}

\section{Introduction}
\label{intro}
Starting from 1980, the polar regions have been warming approximately twice as fast as the mid-latitudes in the Northern hemisphere, a phenomenon known as Arctic warming amplification. Model simulations (the 40 member CCSM3 ensemble) imply that this trend will continue in the future due to a robust global warming signal, with an average warming of 2–3 standard deviations over the Northern Hemisphere continents \citep{Wallace_et_al.:2014}. The signal-to-noise ratio is relatively small at higher latitudes and more significant in the tropics. Despite its well-established presence, there is still a debate about the leading causes of Arctic warming amplification. Different models suggest that sea-ice loss, lapse-rate feedback, or increasing downwelling radiation at the surface could be the main contributor \citep{stuecker2018polar}. Regardless of the causes, it is not clear whether this faster Arctic warming impacts the large-scale circulation and, if so, what the effects of such changes are on extreme events. \cite{francis2015evidence} found a robust link between Arctic amplification and a slower wavier jet stream. This slow down in the jet stream propagation, in turn, would impact extreme weather events---such as heatwaves, heavy downpours, and hurricanes---by increasing their duration and frequency. Other studies, however, dispute this robust evidence linking the polar warming amplification to increasing north–south wave amplitude and slower propagation speed \citep{trenberth2014seasonal,cohen2014recent, Blackport_and_Screen2020}.

The two major wave features that have been connected to extreme events and polar warming are quasi-resonant amplification of Rossby waves trapped in a waveguide and blocking \citep{petoukhov2013quasiresonant}, a persistent pressure anomaly that prevents the usual zonal propagation of atmospheric perturbations \citep{Benzi_et_al.:86}. The latter is connected to Arctic warming assuming that a reduced temperature gradient leads to a slow down of the background westerlies and the drifting Rossby waves, and hence blocking and extreme events become more likely. 
A recent study by \cite{Riboldi_et_al.:2020} confirms the link between low phase speeds and extreme temperature events but also indicates that Arctic amplification did not significantly modulate phase speed variability in recent decades. Blocking has been studied experimentally as well but in connection with zonal asymmetries either in the form of topography \citep{Weeks_et_al.:1997} or with a local heat source in a rotating annulus \citep{Marshall_and_Read:2020}.

The complex dynamics of the atmosphere and the multiple influences and feedbacks pose severe difficulties in finding a final answer to the effects of Arctic amplification on the mid-latitude weather patterns \citep{overland2016nonlinear}. Laboratory experiments enable us to isolate the two key elements of mid-latitude variability from any other feedback process. Such isolation is fundamental in understanding the dynamical cause-effect connection. Furthermore, laboratory experiments are repeatable and can simulate very long time series, providing a statistically significant data set. 
Hence,  we propose an experimental study to complement the widely used observational data. 
For this study, we use a differentially heated rotating annulus \citep{vincze2015benchmarking}, which has been widely used to model the atmospheric jet stream \citep{hide1958experimental} and nonlinear interactions between Rossby wave trains and the mean flow \citep{Read_review}.

Over many years, these laboratory experiments have played a prominent role in geophysical fluid dynamics and climate studies. In the review article by \cite{Vincze_and_Janosi:2016}, several examples are given, such as the investigation of asymmetries of atmospheric temperature fluctuations and experiments on interdecadal climate variability. The emerging scenario reveals that local variability, e.g. in Western Europe or North America, has been increasing in the past 40 years. \cite{Vincze_et_al.:2017} investigated the nature of connections between external forcing and climate variability conceptually using a laboratory experiment subject to continuously decreasing ‘pole-to-equator’ temperature contrast $\Delta T$. Finally, \cite{rodda2020transition} recently demonstrated the potential for using laboratory data to study multiple-scale interactions and explain even mesoscale atmospheric processes. Their study reveals that frequency spectra from the differentially heated annulus experiment are comparable to the power spectra from atmospheric field observations.

The paper is structured as follows. In section 3, we briefly describe the experiment, and in section 4, we give insight into typical flow regimes of the annulus experiment. We further study the impact of polar warming on the wave train structure and drift speed. We then inspect some of these features in NCEP reanalysis data in section 5. Section 6 is the core of this paper where we present a global and regional study of the extreme event frequency from experimental data. The findings are compared with results from section 5. We also discuss probability distributions under polar warming. Finally, in section 7, we offer our concluding remarks.

\section{Experimental apparatus and measurements}
\begin{figure}
\centering
\includegraphics[width=0.5\textwidth]{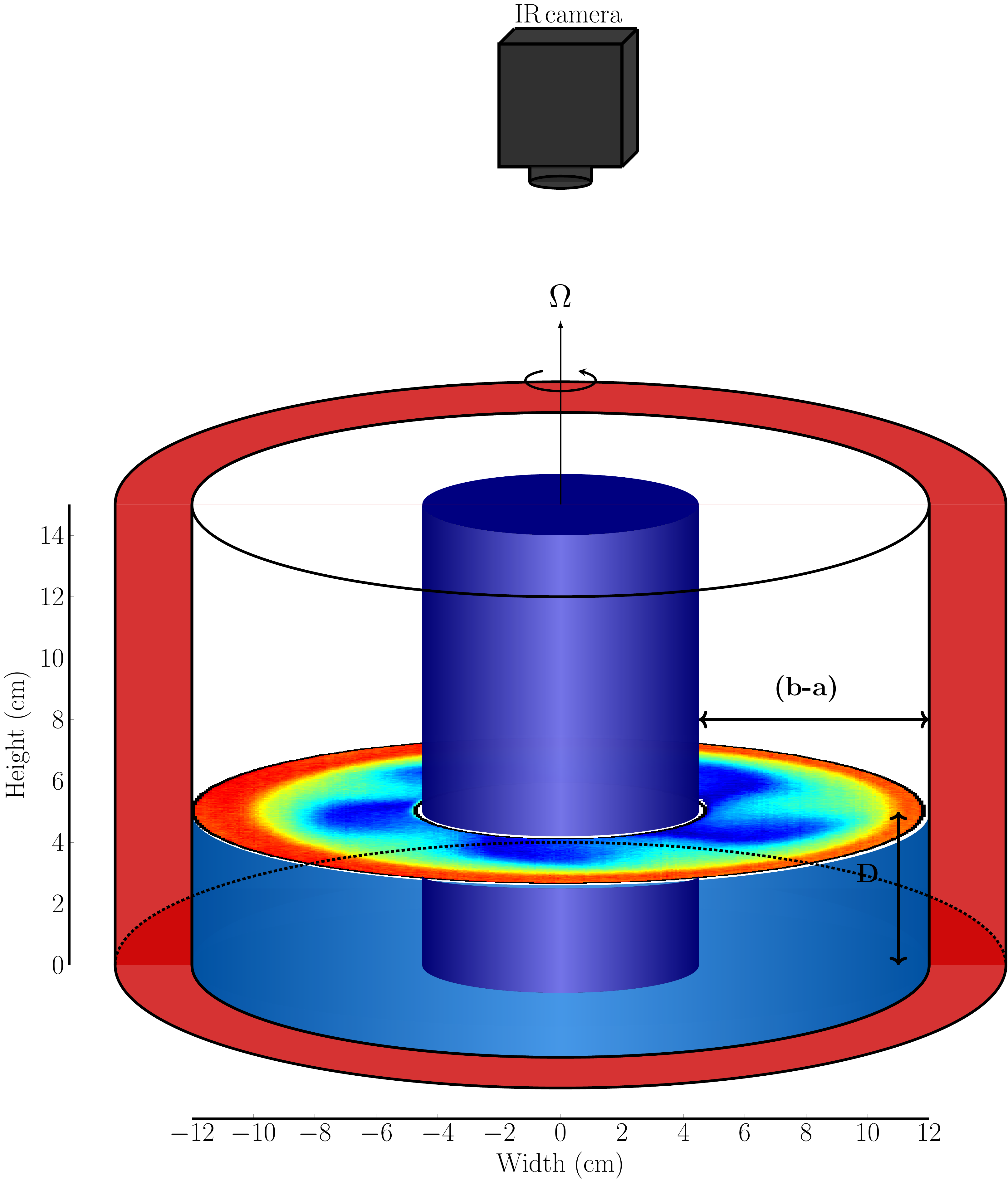}
\caption{Schematic drawing of the experimental apparatus. The inner cylinder is cooled using a thermostat. The outer wall is heated by fluid surrounding the outer wall. The surface temperature of the fluid in the gap between the walls is measured by an infrared camera aligned with the axis of rotation.}
\label{fig:sketchexp}
\end{figure}
The experiments presented in this paper are run with a differentially heated rotating annulus at the BTU Cottbus-Senftenberg laboratories (see \cite{von2005experiments} for more details about the experimental apparatus). The experimental setup, sketched in figure \ref{fig:sketchexp}, consists of a cylindrical tank divided into three concentric rings. The inner cylinder has a radius $a=4.5$ cm and is made of anodised aluminium; the outer ring is separated from the middle cavity by a borosilicate glass wall fixed at radial distance $b = 12$ cm. The three rings are filled with de-ionised water. Heating wires in the outer ring warm up the water, whilst the water filling the inner cylinder is cooled by an external thermostat connected to the experiment. Both the warm and cold reservoirs are kept at a constant temperature throughout the entire duration of each experimental run. The mid-gap water filled up to the height $D=5$ cm is subjected to a radial temperature difference imposed at the boundaries by the two thermally conducting walls. The tank is mounted on a turntable with a constant rotation rate around its vertical symmetry axis. Because of the combined effect of the radial temperature difference produced by the two thermal baths and the rotation of the tank, the baroclinic instability can develop in the middle gap giving rise to baroclinic waves, which are analogous to the atmospheric mid-latitude jet stream.
\begin{figure}
\centering
\includegraphics[width=0.5\textwidth]{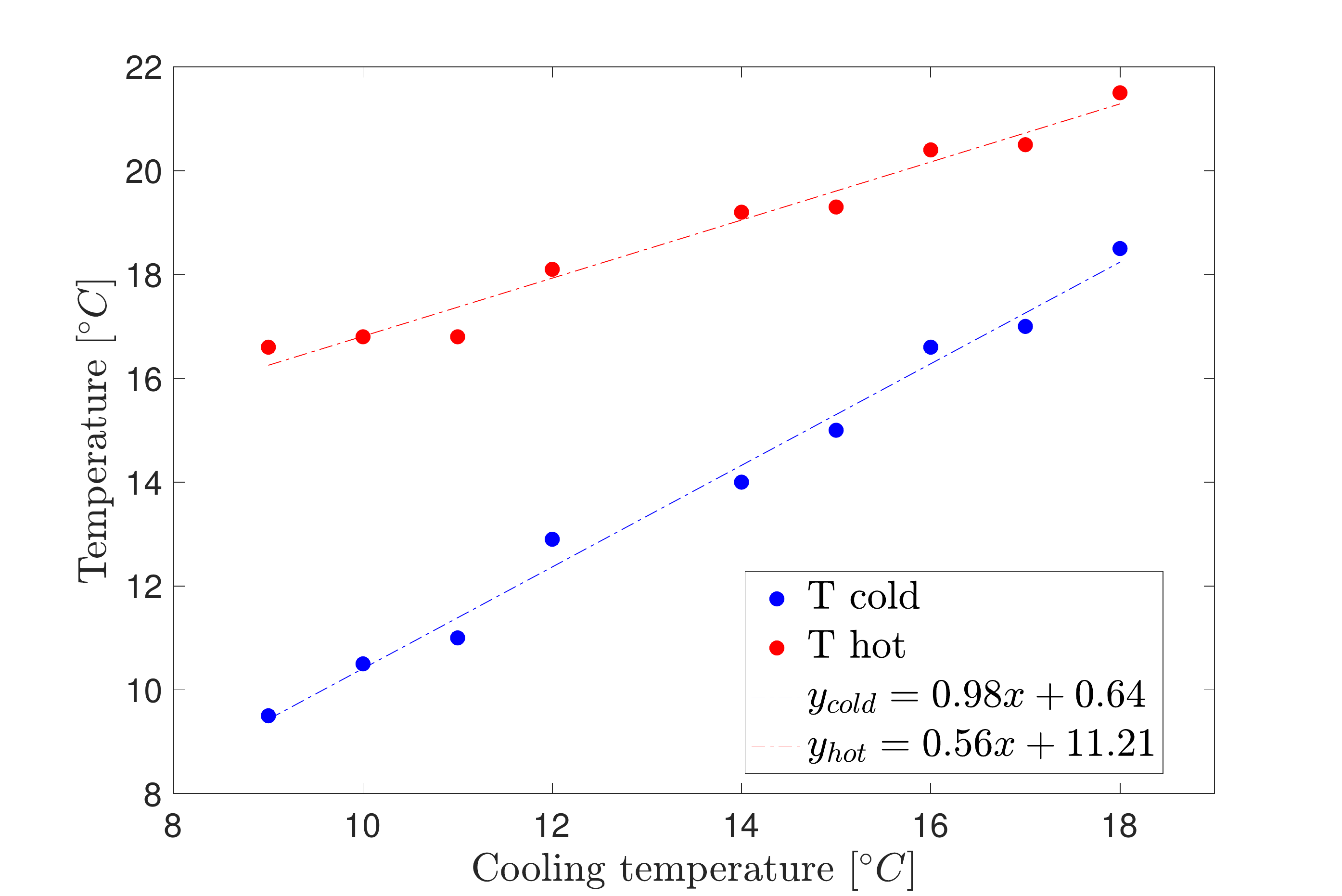}
\caption{Temperature trends in the cold inner and hot outer rings as a function of the imposed temperature of the cooling thermostat. The cold cylinder warms up 1.75 times more than the hot ring in response to the cooling temperature increase.}
\label{fig:linearCoolMean}
\end{figure}
Our experimental data set consists of nine runs that only differ in the cooling temperature, spanning from 9 $^{\circ}$C to 18 $^{\circ}$C. All the other parameters are kept invariant throughout the different runs. Each experiment is run as described in the following. After setting the temperature in the cold and hot baths, the experiment is run for two hours until a constant azimuthal temperature difference $\Delta T$ is reached. Then the rotation rate $\Omega$ is set to 8 rpm. After the spin-up time (approximately two hours), the data are collected for seven hours, corresponding to ca. 3000 revolutions of the tank. Given that each revolution corresponds to one terrestrial day, our experiment simulates nine different cases with different pole-to-equator temperature differences lasting each for little more than eight years in the lab-Earth analogy. \\

The surface temperature is measured with an Infrared Camera, IR camera in short (Jenoptik camera module IR-TCM 640, with thermal sensitivity of 0.01 K and image resolution $640 \times 480$ pixels). The IR camera is fixed in the laboratory reference system and mounted at the top of the experiment (see figure \ref{fig:sketchexp}). The IR camera outputs are time series of the entire annulus 2D surface field measured once at each tank rotation, which corresponds to a sampling interval $dt=7.5$ s. The advantage of having the 2D field is that some characteristics, such as dominant wave patterns, drift speed, and changes in the structure, can be detected. Furthermore, the spatial and temporal resolution $400 \times 400 \times 3000$ (pixels and times, $x,y,t$), which gives a reliable statistics for studying extreme events.

Sensors are placed in the outer warm and inner cold ring to measure the temperature in the two thermal baths. Figure \ref{fig:linearCoolMean} and table \ref{tab:Truns-IR} show the mean temperature measured by these sensors as a function of the temperature set in the cooling basin of the thermostat for each experimental run. In figure \ref{fig:linearCoolMean}, one can easily notice that both the cold and hot baths warm up due to the increased temperature in the cold thermostat. But the cold ring warming is much higher than the increase in the hot ring. This diverse response to the change in the cooling thermostat reproduces in the experiment a warming scenario similar to Arctic amplification. In the rest of the paper we use the more general term `polar warming' for this scenario. 

\begin{figure*}
\centering
\includegraphics[width=0.8\textwidth ]{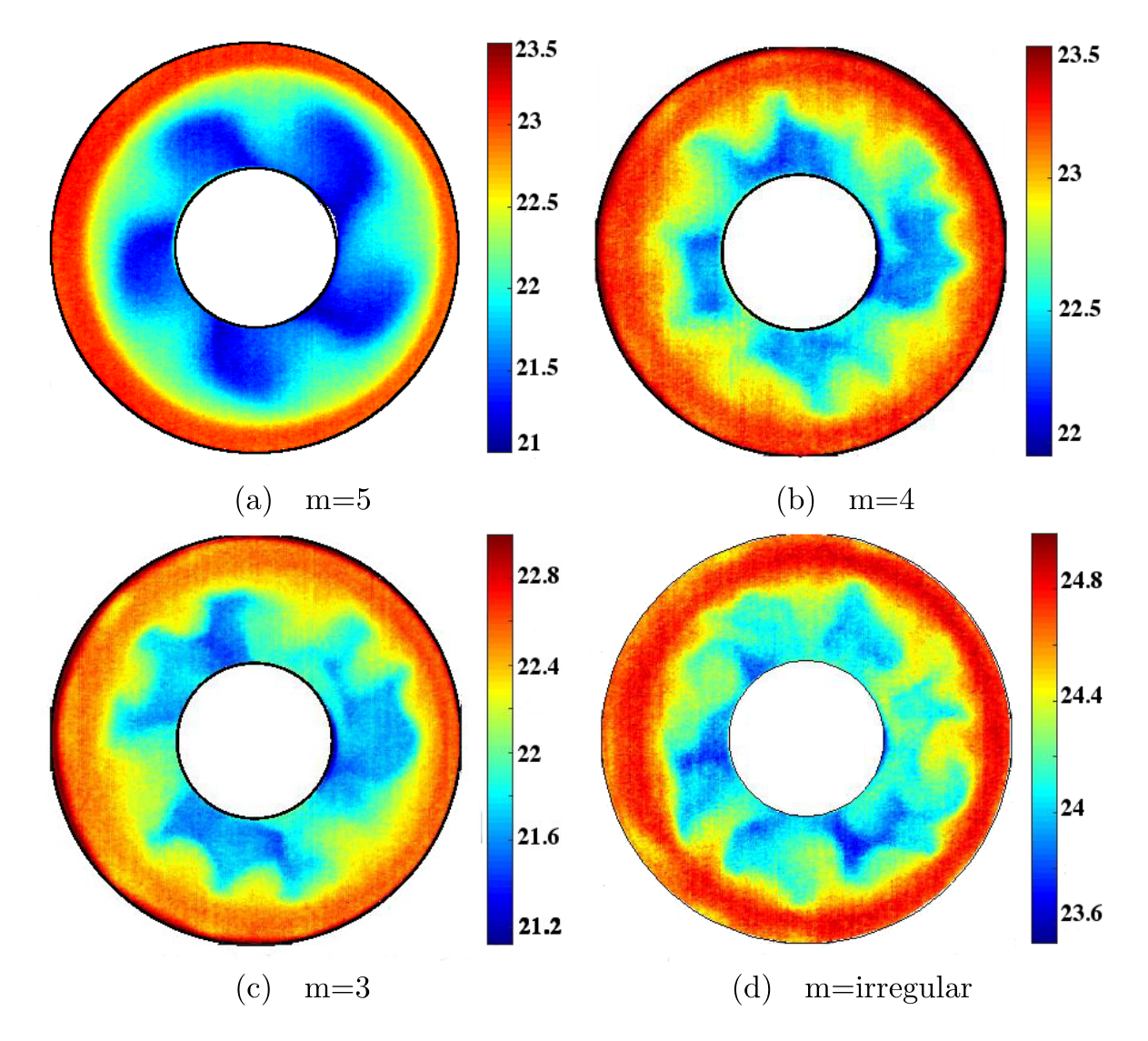}
 \caption{Flow regimes at $\Delta T=7.1 K$ (a), $\Delta T=5.2 K$ (b) and (c), and $\Delta T= 3 K$ (d). The flow becomes more irregular for decreasing $\Delta T$.}
\label{fig:IR-cooling10EOF}
 \end{figure*}
%
\section{Flow regimes}
\begin{table}
\caption{Overview of the experimental runs. $T_C$ is the temperature of the cooling thermostat. $T_{\rm{cold}}$ and $T_{\rm{warm}}$ are the measured temperatures in the inner and outer ring respectively. $\overline{T}=(T_{\rm{warm}}+T_{\rm{cold}})/2$ is the mean temperature. $\Delta T=T_{\rm{warm}}-T_{\rm{cold}}$ is the radial temperature difference. $m$ is the azimuthal wavenumber of the baroclinic wave. The rotation rate is $\Omega=8$ rpm for all runs. $Ro_{T}$ is the thermal Rossby number (\ref{eq:RoT}). The Taylor number (\ref{eq:Ta}) is $Ta=1.32 \times 10^{8}$.}
\label{tab:Truns-IR}
\centering
\begin{tabular}{c | c c c c c c c}
\hline\noalign{\smallskip}
Name & $T_C$ & $T_{\rm{cold}}$ & $T_{\rm{warm}}$ & $\overline{T}$ & $\Delta T$ & $m$ &$Ro_{T}$\\
\noalign{\smallskip}\hline\noalign{\smallskip}
C18 & 18 & 18.5 & 21.5 & 20 & 3 & I & 0.019\\
C17 & 17 & 17 & 20.5 & 18.7 & 3.1 & I-3 & 0.020\\
C16 & 16 & 16.6 & 20.4 & 18.5 & 3.8 & I-3 & 0.024\\
C15 & 15 & 15 & 19.3 & 17.1 & 4.5 & 4-3 & 0.029\\
C14 & 14 & 14 & 19.2 & 16.6 & 4.8 & 4-3 & 0.034\\
C12 & 12 & 12.9 & 18.1 & 15.5 & 5.2 & 4-3 & 0.034\\
C11 & 11 & 11 & 16.8 & 13.9 & 5.8 & 4-3 & 0.037\\
C10 & 10 & 10.5 & 16.8 & 13.6 & 6.3 & 4-3 & 0.040\\
C9 & 9 & 9.5 & 16.6 & 13 & 7.1& 5-4 & 0.046\\
\noalign{\smallskip}\hline
\end{tabular}
\end{table}
This section discusses the dependency of the global flow structures, called flow regimes, on the radial temperature difference $\Delta T$. Understanding of how changes in $\Delta T$ impact the spatio-temporal flow features is a necessary first step for examining the effects on the distribution of extreme events. 

Two parameters control the flow behaviour in a differentially heated rotating annulus \citep{hide1975sloping}: the Taylor number
\begin{equation}
    Ta = \frac{4 \Omega^{2}(b-a)^{5}}{\nu^{2}D},
    \label{eq:Ta}
\end{equation}
and the thermal Rossby number
\begin{equation}
    Ro_T=\frac{gD\alpha\Delta T}{f^{2}(b-a)^{2}}.
    \label{eq:RoT}
\end{equation}
Here, $g$ is gravity, $\alpha=0.207 \times 10^{-3}$ K$^{-1}$ the volumetric thermal expansion coefficient, $\nu=1.004 \times 10^{-6}$ m$^2$s$^{-1}$ the kinematic viscosity of water, $D=0.05$ m the fluid depth, $\Omega=8$ rpm the tank rotation, and $b-a=0.075$ m the gap width.
In general, increasing $Ta$ leads to a more turbulent flow. A small $Ro_T$ enhances geostrophy but also forces 2D turbulence. A larger $Ro_T$ leads to more regular wave regimes. Since we kept the tank rotation $\Omega=8$ rpm constant, the Taylor number, $Ta=1.32 \times 10^{8}$, is constant as well and $Ro_T \sim \Delta T$.

Table \ref{tab:Truns-IR} lists the experimental runs and their parameters. $T_C$ is the temperature of the cooling thermostat. $T_{\rm{cold}}$ and $T_{\rm{warm}}$ are the mean temperatures measured by sensors placed in the inner cold cylinder and warm outer ring, respectively, once a constant temperature difference $\Delta T$ is reached. Due to $T_C$ rising, the mean temperature in the gap $\overline{T}$ (resulting from a thermal equilibrium between the boundary walls) increases, even though the heating power supply to the outer ring is constant. $T_{\rm{cold}}$ and $T_{\rm{warm}}$ are plotted in figure \ref{fig:linearCoolMean} as a function of $T_C$. It can be noticed that $T_{\rm{cold}}$ (in blue) increases 1.75 times more than $T_{\rm{warm}}$ (in red in figure \ref{fig:linearCoolMean}). This temperature change is well suited to mimic the polar warming effect observed in the atmosphere experimentally.
%
\begin{figure*}
\centering
\includegraphics[width=0.9\textwidth ]{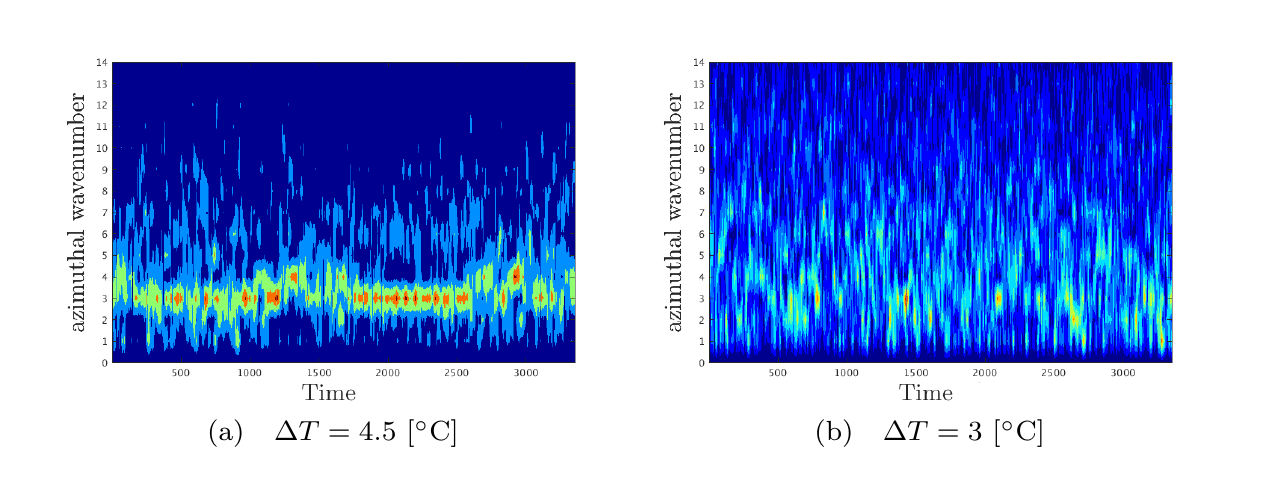}
 \caption{Azimuthal wavenumber evolution over time for experiments C15 (a) and C18 (b). The colormap represents the Fourier transform of temperature data sampled at constant radius $r_0=0.063$ m. Larger $\Delta T$ are associated to flows with a dominant wavenumber where sporadic transitions occur over time (a). When $\Delta T$ is decreased the flow is irregular as a result of a superposition of wavenumbers (b).}
\label{fig:FFTs}
 \end{figure*}
\subsection{Wavenumber transitions}
It is enlightening to study the flow regimes in the azimuthal wavenumber space. 
The column noted with $m$ in table \ref{tab:Truns-IR} indicates the dominant azimuthal wave number observed during the data acquisition. Roughly speaking, decreasing $\Delta T$ leads to smaller dominant wavenumbers and a more irregular flow. Figure \ref{fig:IR-cooling10EOF} displays surface temperature snapshots taken with the IR camera during different experimental runs and gives examples of flow regimes. Run C9 (figure \ref{fig:IR-cooling10EOF} (a)) has a regular flow regime with a regularly shaped baroclinic wave $m=5$. All the other experiments, with smaller $\Delta T$, exhibit more irregular patterns with wavenumbers $m \le 4$, which spontaneously transition to other $m$ over time. This transition can be seen in the time evolution of the azimuthal wavenumber, determined by calculating the spatial Fourier transform of the temperature data measured with a sampling rate of $\Delta t=3.75$ s along a constant radius $r_0=0.063$ m. In experiment C15 (figure \ref{fig:FFTs} (a)), the flow is for most of the time in a state with $m=3$ showing sporadic transitions to $m=4$. In contrast, in experiment C18 (figure \ref{fig:FFTs} (b)) the flow is in a noticeably more turbulent state, with a broad spectrum of excited wavenumbers. In the latter regime, it is challenging to identify a dominant azimuthal wavenumber $m$ and, therefore, we refer to this state as irregular (indicated by $I$ in table \ref{tab:Truns-IR}).

\subsection{Drift speed}
Combining $U_T=Ro_T f L/4$ with (\ref{eq:RoT}), the thermal wind balance gives a first approximation of the background zonal flow
\begin{equation} \label{wind}
    U_T = \frac{g \alpha D \Delta T}{f(b-a)}.
    \label{eq:thermal}
\end{equation}
The linear theory by \citet{eady1949long} predicts that the drift speed of an unstable baroclinic wave in a zonal shear flow $U=U_0 z$ ($U_0$ constant) of depth $D$ is $c=D U_0/2$. It follows that baroclinic Eady waves are non-dispersive, and each wave mode drifts with the same speed. Previous experimental studies (\cite{fein1973experimental, vincze2015benchmarking}) have observed a reasonable agreement between the measured drift speed of the dominant mode and a more general power law equation
\begin{equation}
c=B(\alpha \Delta T/f)^{\zeta}, \label{eq:cm}
\end{equation}
with $B=gD/(b-a)$ a constant depending on the experimental configuration and the exponent $\zeta$ an empirical parameter. Note that for $\zeta=1$, we obtain the thermal wind $U_T$ given by (\ref{wind}). However, the real flow is not simply a linear function of $z$, particularly in experiments with a free surface. Moreover, due to the lateral walls, the mean flow shows also a shear in the radial direction. Finally, rotation and the curvature of the side walls lead to a slight deformation of the free surface. These effects taken together introduce a weak dispersion to the Eady waves. Low-frequency vacillations might be a consequence of this dispersion even without nonlinear interactions, as discussed in detail by \cite{Harlander_et_al.:2011}. Therefore, the drift speed of the baroclinic wavefront is challenging to predict, especially in regimes at which the dominant wavenumber is changing over time (see figure \ref{fig:FFTs}).    
\begin{figure}
    \centering
    \includegraphics[width=0.5\textwidth]{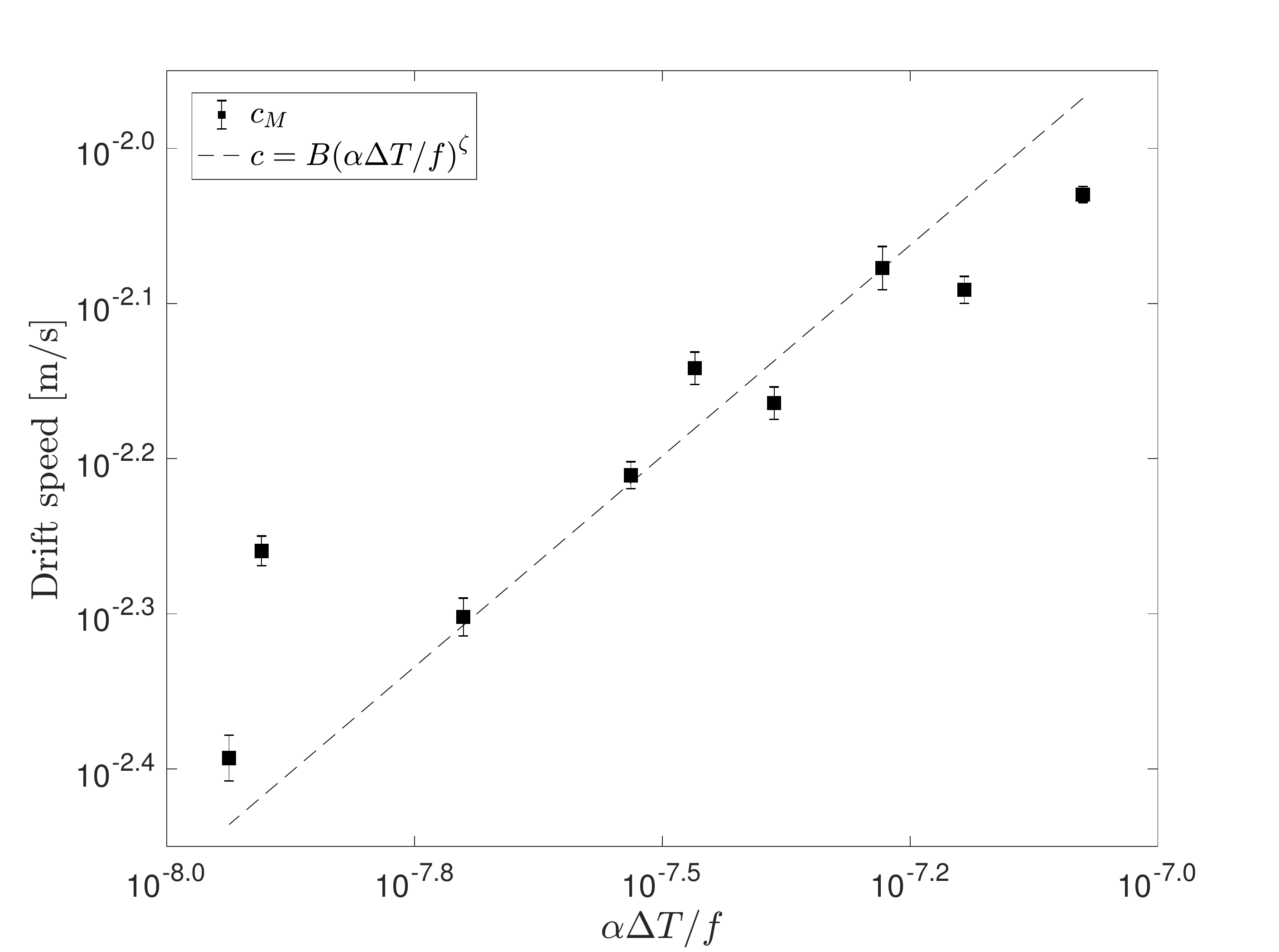}
    \caption{Drift speed $c_M$ of the baroclinic wavefront for decreasing $\Delta T$ calculated from $\theta$-$t$ diagrams at a fixed radius $R_d = 0.087$ m. The dashed line $c$ is calculated from (\ref{eq:cm}) with $\zeta=0.55$. The decrease in the radial $\Delta T$ affects the baroclinic flow by slowing it down.}
    \label{fig:driftspeed}
\end{figure}
In general, the drift speed $c_M$ can be defined as: 
\begin{equation}
    c_M=\frac{1}{m} \frac{d \theta}{dt}=\frac{1}{m} \omega=\frac{\lambda}{T_p}\approx {U_T},
    \label{eq:drift}
\end{equation}
where $\omega=d \theta/dt, \lambda=2 \pi/m, T_p=2 \pi/\omega$ are wave frequency, wavelength, and period, respectively. 
Therefore, if the measured temperature at a fixed radius $R_d$ is plotted as a function of $\theta$ (azimuthal angle) and time, then the drift speed can be graphically computed. This is done by taking $\Delta \theta=\lambda/R_d$ and $T_p$ as vertical and horizontal distances of the wave crests in the plot. For each $\Delta T$, the drift speed is calculated with the above-explained method for ten time intervals spanning over the total measurement duration. Then the mean and the standard deviation are calculated. 
The mean drift speed $c_M$ with the associated error is plotted in figure \ref{fig:driftspeed} as a function of $\alpha \Delta T/f$. It is evident that decreasing $\Delta T$ slows down the drift speed of the baroclinic waves. The dashed line in figure \ref{fig:driftspeed} depicts the general power-law equation (\ref{eq:cm}), where we calculated the coefficient $\zeta=0.55$  by fitting the data. 
 
Differences between the theory and the real flow notwithstanding, the thermal wind equation is close to the experimental measurements, particularly for the flow regimes in which a more regular $m=4$ wave is observed (the intermediate values of $\Delta T$ in runs C11-C15, see table \ref{tab:Truns-IR}). Furthermore, the reduced slope for the experimental data is consistent with the occurrence of dispersion. Compared with Eady waves, dispersive Rossby waves are always slower than the mean background flow. This holds in particular for long waves. 

Thus, from the drift speed results we can conclude that the effect of decreasing the radial temperature difference is to slow down the baroclinic front. Hence, the propagation speed of extreme events embedded into the wave trains, e.g. in the form of exceptional large meanders, will be slowed down, and the events in real atmospheric flows might unfold their local destructive potential over a more extended period.  
\section{Extreme events in NCEP data}
\begin{figure}
\centering
\includegraphics[width=0.5\textwidth]{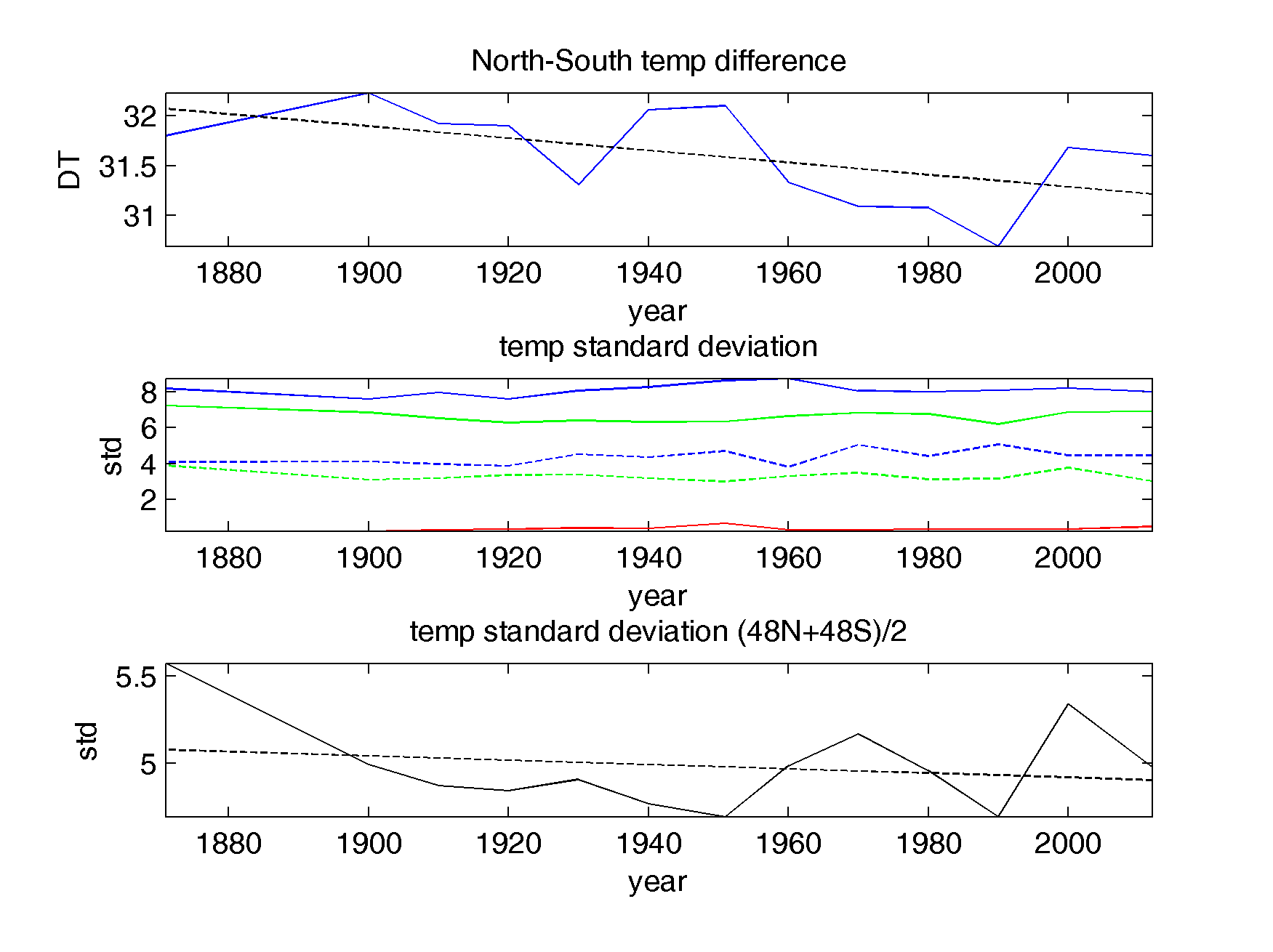}
\caption{NCEP reanalysis data of 500hPa temperature. Upper figure: $\Delta T$ trend from 1871 to 2012. The linear regression (dashed line) shows a clear trend towards smaller north-south temperature differences ($T(2012)-T(1871)=-0.86K$). Center figure: zonal mean temperature standard deviations for $88^{\circ}N$, $88^{\circ}S$ (blue solid, blue dashed), $48^{\circ}N$, $48^{\circ}S$ (green solid, green dashed), and equatorial (red). Bottom figure: mid-latitude zonal mean temperature standard deviation of ($(T(48^{\circ}N)+T(48^{\circ}S))/2$) (solid line) and linear regression (dashed line). There is a weak trend towards decreasing standard deviation($\sigma_T(2012)-\sigma_T(1871)=-0.18K$). }
\label{fig:NCEP_std500}
\end{figure}
Before we start analysing the lab data for changes of variability and extreme events as a function of the North-South temperature gradient, it is instructive to inspect some of these features in the National Centers for Environmental Prediction (NCEP) reanalysis data \citep{Kalnay_etal:1996}. 
In contrast to operational counterparts, the reanalysis data do not suffer from inhomogeneities introduced by changes in the data assimilation system. In this respect, they are better suited for studying variability and climate trends than data based on individual instrumental records or climate-model simulations \citep{Uppala:2008}.
Moreover, reanalysis data cover historical data as well. We use two temperature data sets from the collection ``NOAA-CIRES 20th Century Reanalysis, version 2, Daily Averages'', covering the period from 1871 to 2012. The first set is the daily ensemble mean pressure level data (1000hPa to 10hPa), from which we extracted just the 500hPa level. The second set is the daily ensemble mean tropopause data.
\begin{figure*}
\centering
\includegraphics[width=0.8\textwidth]{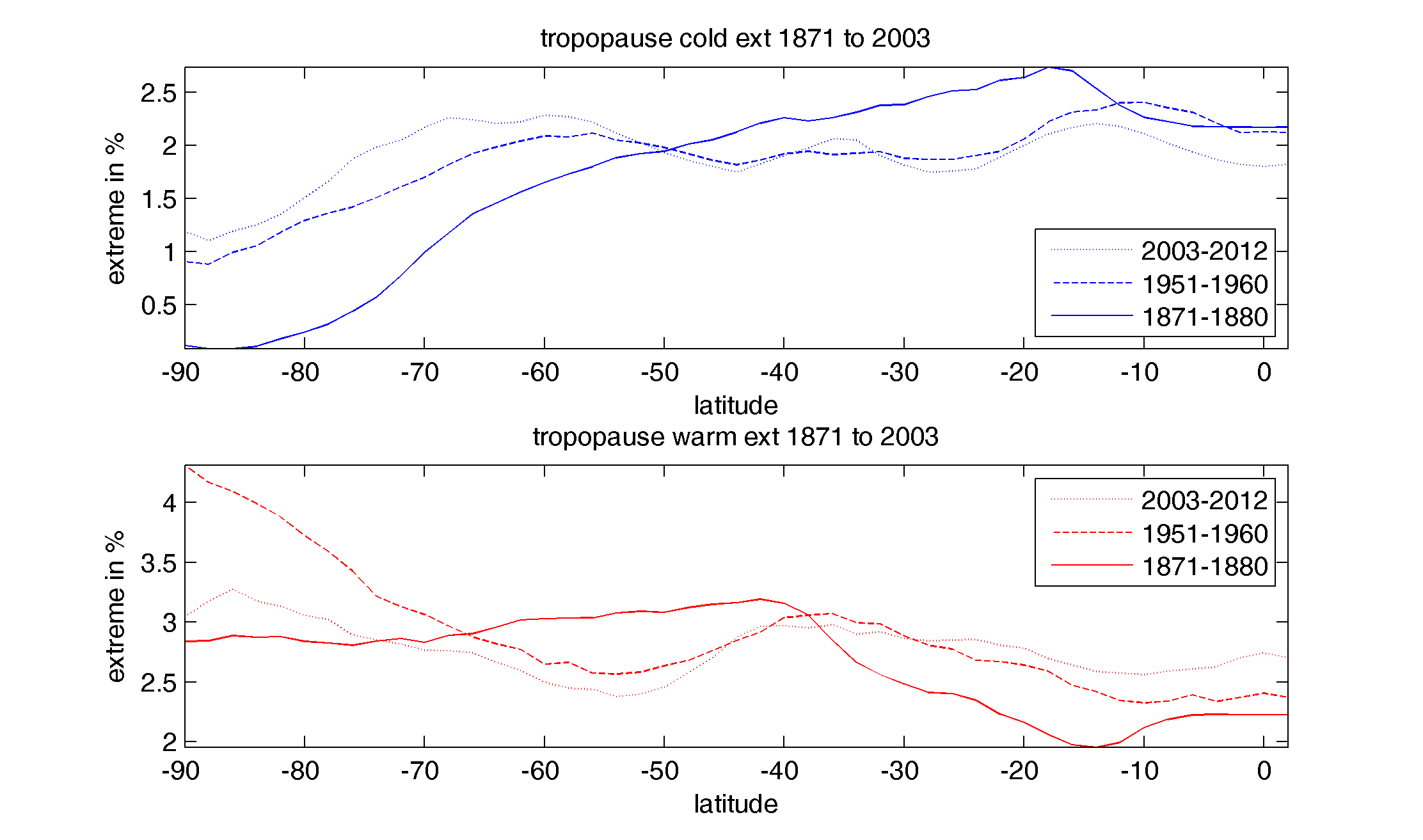}
\caption{NCEP reanalysis data of tropopause temperature. Tropopause data are since the frequency spectra of the baroclinic wave experiment looks most similar to tropopause spectra \citep{rodda2020transition}. Pole-to-equator extreme events trend. The upper and bottom plots depict the cold and hot extreme events, respectively. The lines are: solid for the period 1871-1880, dashed for 1951-1960, dotted for 2003-2012. The seasonal cycle has been removed from the data. The extreme values have been taken from the Pacific sector (weak land sea contrast) and the frequencies from the southern and northern hemisphere have been averaged for any latitudinal circle.}
\label{fig:NCEP1}
\end{figure*}
We start by considering the gradient and the standard deviation of the 500hPa temperature from 1871 to 2012. We know from previous studies \citep{Walsh:2014} that over the past 50 years the annual mean temperature above the extratropical Northern Hemisphere has increased about twice the global rate and even stronger with respect to tropical regions. In the literature this is called polar amplification. Polar amplification implies that the north-south temperature gradient has reduced. 

The upper panel of figure \ref{fig:NCEP_std500} shows the trend in the north-south temperature difference ($\Delta T$) taken from the NCEP 500hPa data. This temperature gradient is evaluated from the zonal mean temperature at $88^{\circ}N$ ($88^{\circ}S$) and the equatorial zonal mean temperature. The polar temperature is calculated as $(T(88^{\circ}N)+T(88^{\circ}S))/2$. We see that, as expected from what we said above, the gradient has been weakened in time. The linear regression (dashed line) shows a clear trend towards smaller north-south temperature differences ($T(2012)-T(1871)=-0.86K$). The central figure displays the time evolution of the standard deviation $\sigma_T$ at different latitudes (blue solid, $88^{\circ}N$; blue dashed $88^{\circ}S$; green solid, $48^{\circ}N$; green dashed, $48^{\circ}S$; red, equator). The Northern Hemisphere exhibits a more significant variability, which might be related to a stronger land-sea contrast. For the mid-latitude time series, a small trend can be seen (green lines). In the bottom figure, we highlight this trend by plotting just the mid-latitude zonal mean standard deviation of $(T(48^{\circ}N)+T(48^{\circ}S))/2$ (solid line) and adding the linear regression (dashed line). Obviously, there is a weak trend towards decreasing standard deviation ($\sigma_T(2012)-\sigma_T(1871)=-0.18K$). Such a trend is consistent with other model data \citep{Rind:1989}. We will see later that these observations are also in qualitative agreement with the results from the laboratory experiment.

To inspect the frequency of extreme events, we focused on NCEP reanalysis data of the tropopause temperature. These data are less prone to the effects from land-sea contrasts and might be closer to the lab experiment. \cite{rodda2020transition} have shown that spectra from tropopause data are comparable to the frequency spectra of the baroclinic wave experiment and we suggest a similar connection with respect to extreme values. For our analysis we highlighted three ten year periods: from 1871 to 1880, from 1951 to 1960, and from 2003 to 2012. The seasonal cycle has been removed from the data. The frequency of extreme values, defined as values larger or smaller than twice the standard deviation, has been calculated for all longitude circles with an increment of $20^{\circ}$. Subsequently, we took the mean of the Southern and Northern hemisphere frequencies and finally we zonally averaged the frequency data. This gives the mean frequency of extreme values as a function of latitude ranging from $-90^{\circ}-0^{\circ}$.

\begin{figure}
\centering
\includegraphics[width=0.5\textwidth]{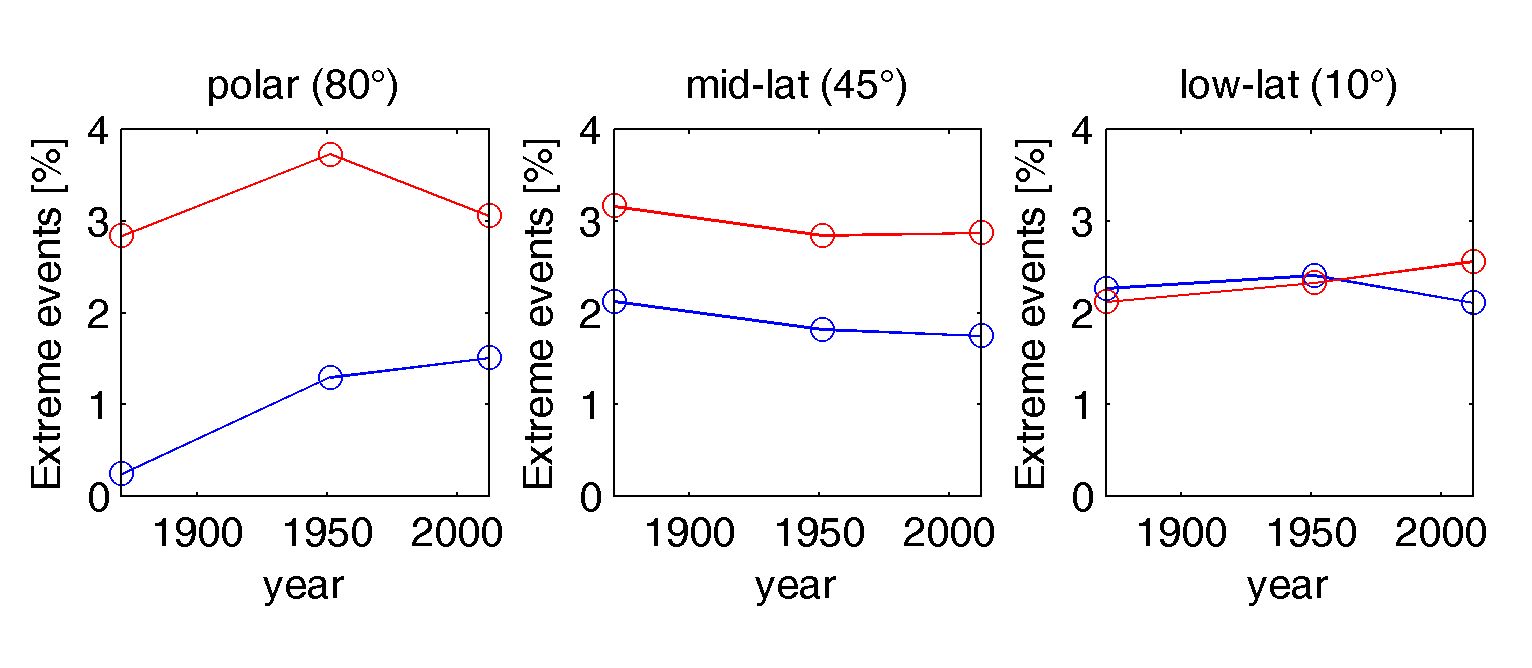}
\caption{NCEP reanalysis data showing the trend of extreme temperatures at different latitudes (left, $80^\circ$; centre, $45^\circ$; right, $10^\circ$). The lines are: solid red for warm events, solid blue for cold events. The seasonal cycle has been removed from the data. The extreme values have been taken from the Pacific sector (weak land sea contrast) and the frequencies from the southern and northern hemisphere have been averaged for any latitudinal circle.}
\label{fig:NCEP2}
\end{figure}

We see from figure \ref{fig:NCEP1} that the number of extremes, broken down to values above (warm) or below (cold) the $2 \sigma$ standard deviation threshold, ranges between $1$ and $4 \%$. For the cold case, the frequency of extreme events is larger at low latitudes. We can see that the period 2003-2012, with the smallest north-south temperature gradient, shows most extremes for polar latitudes but least for tropical latitudes. Note that the experimental data lack of a tropical dynamics (f-plane with a wall as a southern boundary). Hence, one must be careful when comparing low-latitude NCEP data results with experimental data close to the heated outer sidewall. For the warm extremes, we see the highest frequency near the polar region, and again, the periods with smaller north-south temperature gradients show most extremes. 

In figure \ref{fig:NCEP2}, we can see the trend of the extreme events (temperatures above or below the $2\sigma$-value) during the last 140 years. To be exact, we calculated the extreme event frequency for the periods 1871-1880, 1951-1960, and 2003-2012 (see also figure \ref{fig:NCEP1}) and plotted them for the warm (red) and cold (blue) events at three different latitudes. We see a clear trend only for the polar cold events. Furthermore, warm events are more frequent for polar and mid-latitudes.

Finally, we considered the tropopause pacific sector temperature distribution for the three chosen periods and three latitudes, `p' (polar, $80^\circ$), `m' (mid-latitude, $45^\circ$), and `e' (equatorial, $10^\circ$). Since the tropical tropopause is about 5km higher, it is much colder than the mid-latitude and polar tropopause. The north-south temperature gradient is, therefore, positive in contrast to the negative gradient in the troposphere. The impact of climate change on the tropopause dynamics is still under debate \citep{Randel_and_Jensen:2013}. Trends in tropical tropopause temperature result from the combined trends in tropospheric and stratospheric temperatures. Since these trends are of opposite sign, and the sign changes in the tropopause region, it is unclear whether tropopause temperature will increase or decrease \citep{Gettelman:2009}. However, it seems that due to climate change the tropical tropopause will rise (negative pressure trend), and a slight warming trend shows in the tropopause region.

We indeed found from the NCEP data that, in the equatorial region, the tropopause mean temperature is rising from $194$ K to $196$ K from the period 1951-1960 to 2003-2012. At mid-latitudes, the temperature slightly decreases from $214$ K to $213$ K, and the polar tropopause does not change. We further found that the standard deviation increases at polar and mid-latitudes from 1871-1880 to 2003-2012, in contrast to the standard deviation at the 500hPa level (see figure \ref{fig:NCEP_std500}), which slightly decreases with decreasing $\Delta T$. For the polar latitude and mid-latitude in the periods 1951-1960 and 2003-2012, we detected a substantial increase in the frequency of those extreme events showing deviations from the mean larger than $\pm 10$ K. The skewness of the extreme event distributions is positive for polar but negative for equatorial latitudes. Also these findings will later be compared with the laboratory experiment data.

\section{Extreme events in the experimental data}
\label{sec:extremeFreq}
In this section, we investigate the extreme event frequency at different latitudes in an experimental global warming scenario. The extreme events for all experimental runs (see table \ref{tab:Truns-IR}) are compared to evince possible trends when the temperature difference, $\Delta T$, between the pole and the equator diminishes, i.e. in a global warming scenario. The extreme events are defined as temperatures exceeding a defined threshold. We chose such threshold as the standard deviation ($\sigma$) of longitude-temporal series calculated at fixed latitudes. We then call extreme cold/hot events all temperatures such that $T>T_{\textnormal{mean}} \pm 2 \sigma$.
\begin{figure*}
\centering
\includegraphics[width=0.99\linewidth]{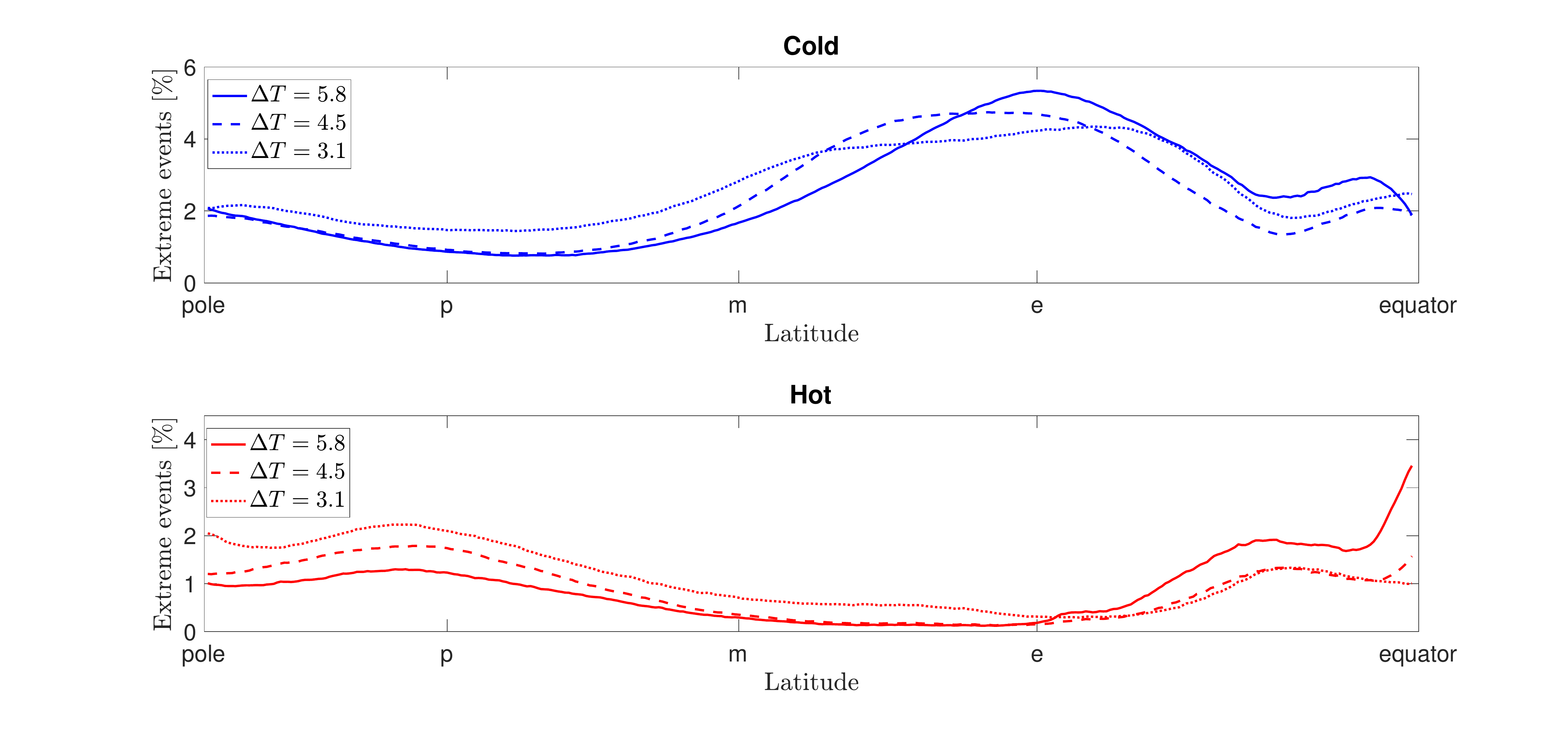}
\caption{Pole-to-equator extreme event trend. The upper plot depicts the cold events, whilst the bottom plot depicts the hot events. The lines are: solid for the experiment at $\Delta T=5.2K$, dashed at $\Delta T=3.8K$, dash-dot line at $\Delta T=3K$. The labels `p', `m', and `e' indicate the latitudes at which the trends shown in figure \ref{fig:trend3lat} are taken (see figure \ref{fig:pme} to visualise their position).}
\label{fig:trend-lat}
\end{figure*}
%
\subsection{Global analysis}
Firstly, we study how extreme events are distributed at different latitudes. For this analysis, temperatures at three fixed latitudes are collected into sets, where each set is constituted by $1.2 \times 10^6$ temperature measurements for different times and longitudes. The standard deviation $\sigma$ is calculated for each set and, successively, the extreme event frequency is calculated as a function of the latitude. Note that, as we will discuss more in detail later, the $\sigma-$threshold is latitude-dependent. The extreme event frequency is defined as the number of times the temperature crosses the set threshold normalised by the total length of the data set (which is, in any case, the same for all the data set considered). The event duration is neglected, i.e. only the number of measurements (``days'') where the temperature exceeds the threshold is counted, without distinguishing whether such days are consecutive or isolated.

\begin{figure}
\centering
\includegraphics[width=0.9\linewidth]{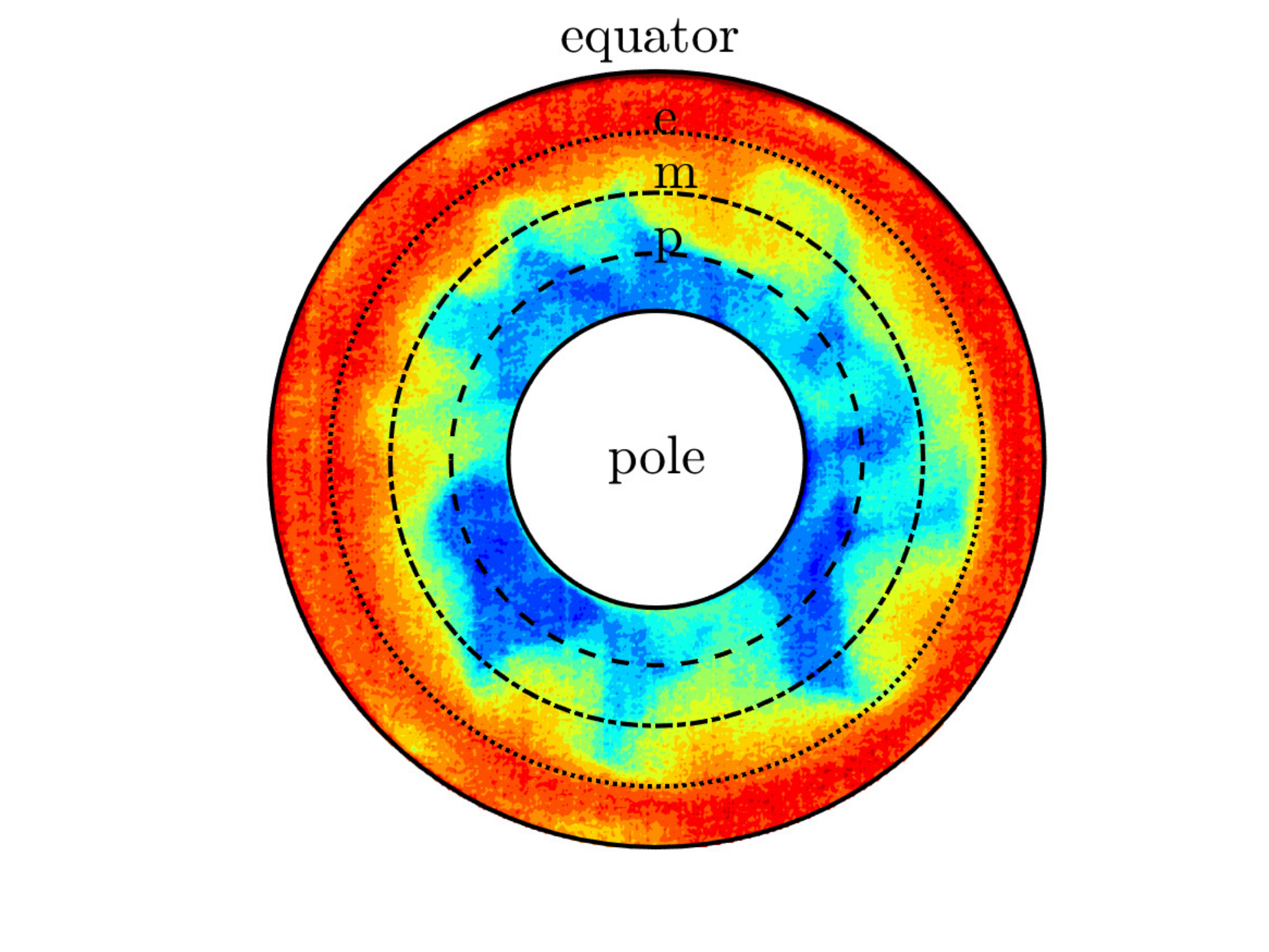}
\caption{Location of the latitudes `p', `m', and `e' with respect to the baroclinic wave for the experiment at $\Delta T=4.5$. }
\label{fig:pme}
\end{figure}

An example of the distributions is reported in figure \ref{fig:trend-lat} for three experiments, namely at $\Delta T= 5.2K$ (solid line) $\Delta T=3.8K$ (dashed line) and $\Delta T=3K$ (dashed-dotted line). For easier visualisation, these three experiments have been chosen as representative for the entire data set. The excluded data show similar results, and therefore the following discussion includes the complete data set. The upper plot presents the extreme cold events $T<T_{\textnormal{mean}}+2 \sigma$, whilst in the lower plot we see the extreme hot events $T>T_{\textnormal{mean}}+2 \sigma$. Three latitudes, labelled as `p',  `m', and `e', are indicated on the $x-$axis for a better readability. The chosen `latitudes' are given in figure \ref{fig:pme}. 

For both plots, we can notice that the three curves have similar shape from the polar latitudes up almost to the point `e', with more frequent extreme events for decreasing $\Delta T$. Yet, the cold and hot extremes exhibit opposite trends: the cold event frequency increases from the pole towards the equator with a maximum just before `e' and then decreases, whilst the extreme hot events behave the opposite.
To some extend, the existence of a local maximum for warm and minimum for cold events at latitude `e' can be understood by looking at figure \ref{fig:pme}, where the latitude `e' (dotted line) marks the limit of the extension of the baroclinic wave cold tongues, covering approximately three quarter of the gap width. The most external region towards the heated wall is unreached by the cold tongues and, therefore, we can expect differences with respect to extreme event frequency in this part of the annulus where we find no baroclinic wave activity. The location of the maximum/minimum in cold/warm event frequency coincides with the maximum latitude to which the baroclinic waves extend and this is a clear indication that the baroclinic wave activity shapes the frequency distribution.

Comparing the experimental data (figure \ref{fig:trend-lat}) with the NCEP data (figure \ref{fig:NCEP1}) we find a striking similarity of the curves from `p' to `e' and from about $-90^{\circ}$ to $-20^{\circ}$, respectively. Cold/warm events recur more at low/high latitudes. Furthermore, the cold events in NCEP and experimental data with large $\Delta T$ are most numerous at low latitudes. The frequency decreases for lessening $\Delta T$. In contrast, for near polar and near equatorial latitudes, cold events are most probable for smaller $\Delta T$. 

It should be noted that the NCEP data (figure \ref{fig:NCEP1}) do not show such distinct extreme event peaks around the near equator region but have a monotonic increase/decrease for cold/hot events instead. This difference can be expected since the appearance of a local peak in the experiment might be due to the mentioned missing tropical dynamics, as previously explained. Note further that the largest $\Delta T$ experiment gives a very regular baroclinic wave (see figure \ref{fig:IR-cooling10EOF} a), rather unrealistic with respect to atmospheric flows. Caution is, therefore, required when comparing the $\Delta T=5.2$ experiment with atmospheric temperature data.

In summary, we suggest that the extreme event spatial frequency distribution is governed by the large-scale baroclinic wave dynamics in the atmosphere as well as in the analog experiment. This explains the strong resemblance between the NCEP and the experimental data.

%
\subsection{Local analysis}
\begin{figure*}
\centering
\includegraphics[width=0.9\linewidth]{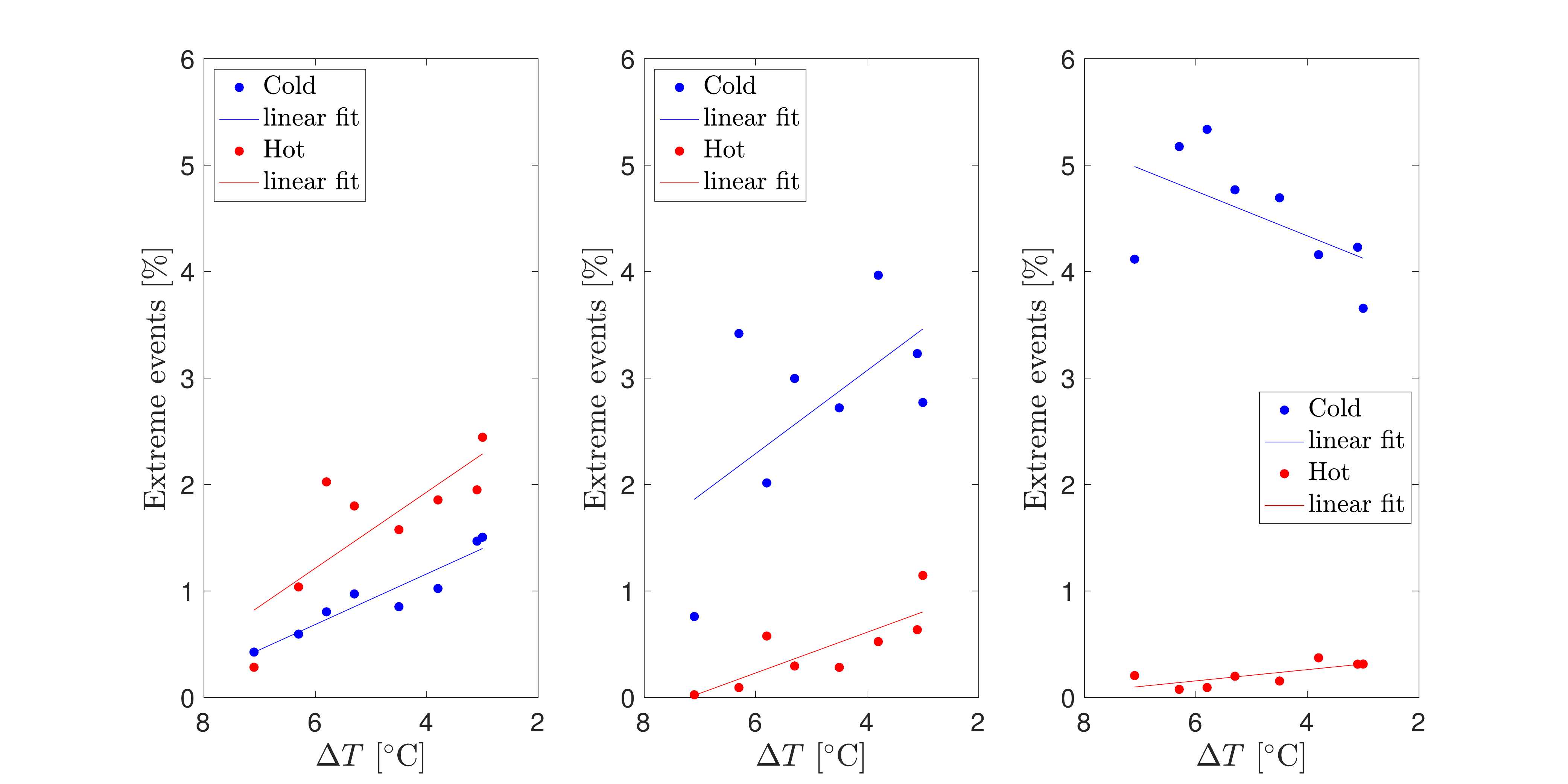}
\caption{Trend of the cold (blue) and hot (red) extreme events with decreasing equator-pole temperature difference. The trends are considered at three latitudes in the plot: polar latitude (left), mid-latitude (middle), and equatorial latitude (right). The three latitudes are indicated in figure \ref{fig:trend-lat} by the labels `p', `m', and `e', respectively.}
\label{fig:trend3lat}
\end{figure*}

We now plot in figure \ref{fig:trend3lat} the relative number of extreme events at the three locations `p', `m', and `e', as a function of $\Delta T$ for the full ensemble (see table \ref{tab:Truns-IR}). The cold events are plotted in blue and the hot in red. What can immediately be noticed by comparing the three plots is that the number of extreme cold and hot events is similar in the region near the pole (left figure), but the extreme cold events become more and more frequent moving towards the equator (middle and right plots). 

The second noticeable feature is that, although there are some outliers, a clear trend characterises the left and middle plots, proving that both cold and hot extreme events increment for decreasing $\Delta T$. Instead, the right plot indicates that the extreme hot events increase, consistently with the left and middle plot, whilst the cold extreme event number decreases. This difference in the cold trends between the polar and mid-latitude regions with the equatorial region can be explained by the fact that the equatorial dynamics is not governed by the baroclinic wave dynamics, and we already suggested that the extreme event distribution is linked to the baroclinic waves. The decrease in cold events is consistent with a possible change in the equatorial extension of the baroclinic wave.

Comparing figure \ref{fig:trend3lat} with figure \ref{fig:NCEP2}, we find that the trend is less clear in the NCEP data. Note that the change in $\Delta T$ is much smaller in the NCEP than the laboratory data. At least, we see a positive trend for polar regions in both data sets as well as an unclear trend for the near tropical regions. Moreover, for polar latitudes the warm events are more likely in the NCEP and the experimental data. For the other latitudes, warm events are more likely in the NCEP data in contrast to the experimental findings.

%
\subsection{Probability distributions}
\begin{figure*}
\centering
\includegraphics[width=\textwidth ]{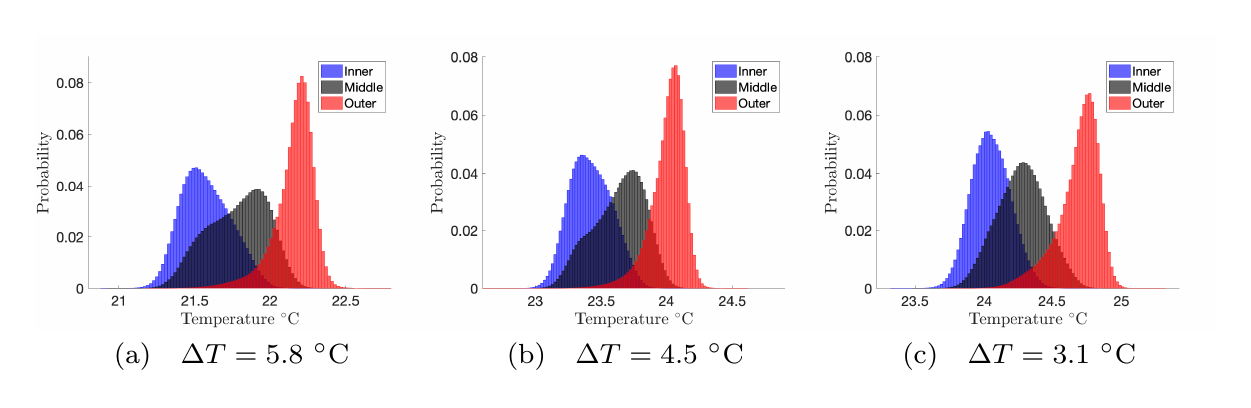}
 \caption{Comparison of probability density temperature distributions for three $\Delta T$. The plots show each one experimental run where the temperature field is divided into three annular regions of equal dimension: the inner most ring (blue), the middle ring (black), and the outer ring (red).}
\label{fig:PSDs}
 \end{figure*}

\begin{figure*}
\centering
\includegraphics[width=0.8\textwidth ]{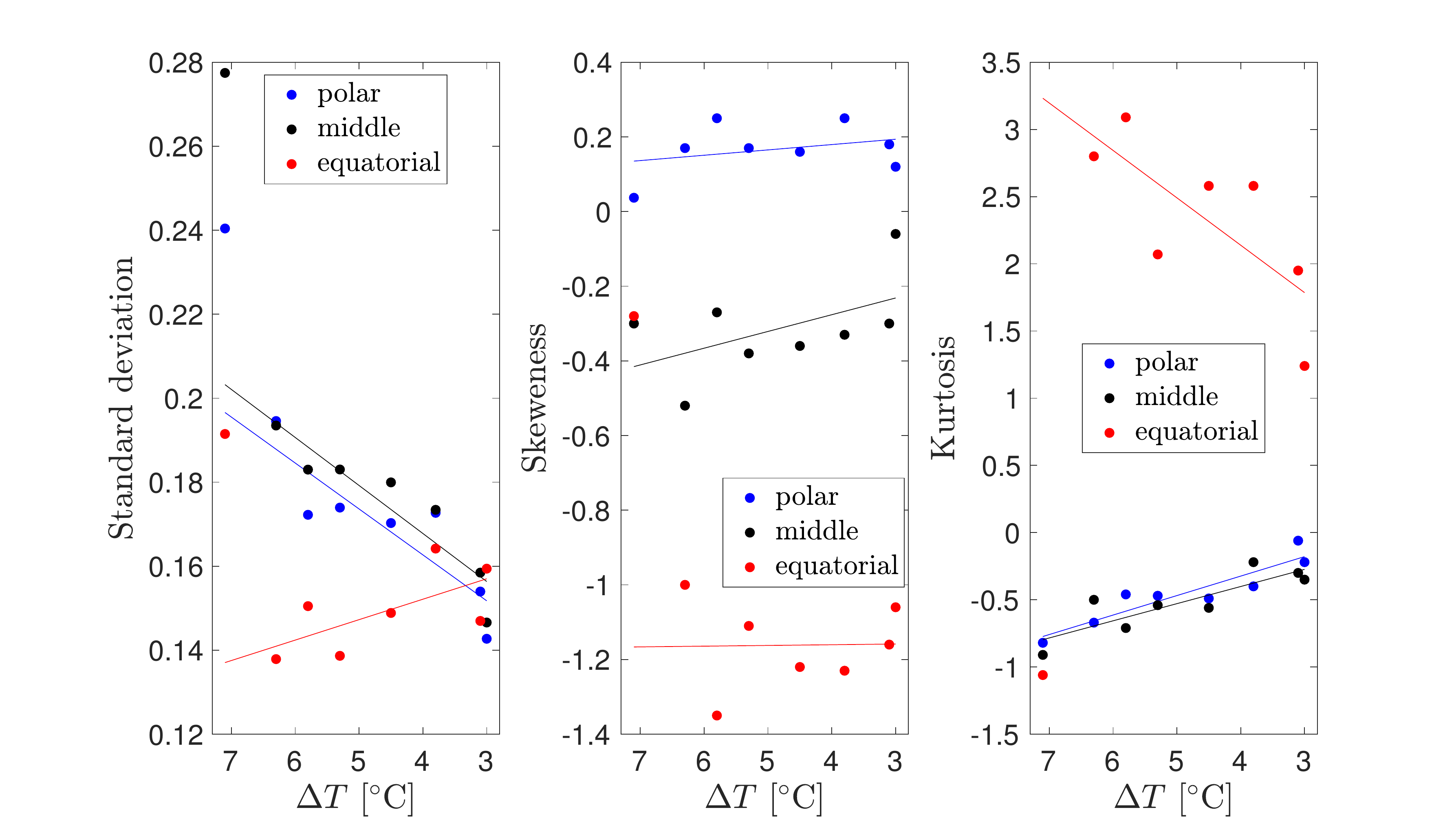}
 \caption{Standard deviation, skeweness and kurtosis trends for decreasing $\Delta T$ for the inner most ring (blue dots), the middle ring (black dots), and the outer ring (red dots). The kurtosis is calculated by subtracting a factor 3, so that a normal distribution is expected to have a value equal to 0. A slight increase in skewness and kurtosis for the black and blue data is related to an enhanced meandering of the jet stream.}
\label{fig:sk-kur}
 \end{figure*}

How do probability distributions of experimental surface temperature data change under `polar warming'? Figure \ref{fig:PSDs}, depicts the distributions for $\Delta T=5.8$ (a), $\Delta T=4.5$ (b), and $\Delta T=3.1$ (c) along the three concentric regions as introduced in figure \ref{fig:pme}: the inner most ring (the polar region is shown in blue), the middle ring (in black) and the outer ring (the equatorial region in red). The major features are rather robust under changes in $\Delta T$: the polar data possess a right (positive) and the equatorial data a left (negative) skew (see also the left plot in figure \ref{fig:sk-kur}).
This behaviour reminds of the results by \cite{Belmonte_and_Libchaber:1996}, who found for turbulent Rayleigh-B{\'e}nard convection that the temperature distribution skewness has a positive value at the cold (top) boundary and becomes more and more negative close to the warm (bottom) boundary.
Similarly, the skewness of the middle and polar regions slightly increases for decreasing $\Delta T$ (figure \ref{fig:sk-kur} (b)) whilst their standard deviations decrease (figure \ref{fig:sk-kur} (a)). In contrast, the skewness and the standard deviation for the outer ring temperature data seem to decrease for a warmer pole. Furthermore, we can notice that the standard deviation has an inverse trend with respect to the extreme events, i.e. $\sigma$ decreases with decreasing $\Delta T$ in the polar and mid-latitude regions (blue and black lines), whilst it increases in the proximity of the equator (red line). This decrease in the standard deviation might indicate that, although we observe more frequent extreme events for smaller $\Delta T$, the temperatures of such extremes are closer to the mean temperature than in experiments for which we have a high standard deviation. Note that the data point corresponding to the highest $\Delta T$ has been ignored to calculate the trends in figure \ref{fig:sk-kur} since it lies far off all the other data points. Maybe most striking is the broadening of the tail for the data at `e' for the small $\Delta T$ value.

It is instructive to compare the temperature distributions with recent findings on distributions of Local Wave Activity (LWA) by \cite{Nakamura_and_Valva:2021}. LWA measures the meridional displacement of quasigeostrophic potential vorticity and is hence a measure for the meandering of the jet stream. Extreme weather events can be attributed to a large meandering. Model data show for both hemispheres a right skew similar to the blue distribution (figure \ref{fig:PSDs}). As shown by the authors, the distributions can be fitted to the Gamma distribution. The skewness results from the nonlinear advection of LWA and, according to the authors, underscores the importance of nonlinearity for the frequency of jet disruptions and associated weather anomalies.

What happens to the LWA when $\Delta T$ is decreased, and can this be related to our findings? A lower baroclinicity, i.e. smaller radial temperature gradient, leads to a slow down of the jet, as shown in figure \ref{fig:driftspeed}. This broadens the distribution of the LWA and should also broaden the temperature distribution. Indeed, this seems to be the case for the temperature data close to the warm outer wall (red distribution in figure \ref{fig:PSDs}) but not for the data close to the inner `p' and centre `m' ring (blue and black distributions in figure \ref{fig:PSDs}). For the latter, the standard deviation is decreasing (figure \ref{fig:sk-kur},left), and the kurtosis is increasing (figure \ref{fig:sk-kur}, right); hence the distributions become narrower and more peaky. This observations fit the fact that a decrease of $\Delta T$ leads to less transient eddies, which narrows the distributions of LWA and temperature. Hence, a subtle interplay between these two effects shapes the distributions under a changing temperature gradient. In the end, these opposite effects might explain the relatively robust distributions under the `climate change scenario' considered. It should be noted that the data from the outer ring (e) might be weaker affected by the baroclinicity changes since, in the experiments, the baroclinic waves only extend to three quarter of the annulus, as we already discussed above.

In summary, our findings on the change of temperature distributions under a reduction of $\Delta T$ fit qualitatively to the results from quasigeostrophic modelling. These are only the first steps, and undoubtedly further research is needed to understand better the coherence between polar warming and probability distributions of suitably defined measures. 

%
%
\section{Conclusions}
We have presented a series of experiments with a differentially heated rotating annulus to model a global warming scenario. The aim of our study is to reproduce the polar warming and study the possible effects on other atmospheric phenomena. For this simple experimental environment, the impact of a reduced pole-to-equator temperature difference onto the mid-latitude large-scale dynamics and consequences for the likelihood and distribution of extreme events could be isolated from various processes that occur and interact in the earth's atmosphere.

In accordance with \cite{francis2015evidence}, we found that the jet stream becomes more irregular as a consequence of warming the pole, making it difficult to identify a clear dominant azimuthal wavenumber. Moreover, the baroclinic wave train drifts slower, which, for the experiment, is a consequence of a slow down of the westerly mean flow for reduced $\Delta T$. Analysing NCEP data, we found in agreement with the experiment more extreme cold/warm events at lower/higher latitudes independently for the time period and temperature differences considered. The experiments show a clear trend towards a larger number of extreme cold and warm events at polar and mid-latitudes. This trend is less clear for the NCEP data. For polar warming, variability is reduced for the reanalysis as well as for the experimental polar and mid-latitude data. 

Due to nonlinear transient effects, temperature data probability distributions are skewed \citep{Nakamura_and_Valva:2021},. The polar and mid-latitude data show a right (positive) and the equatorial data a left (negative) skew. For decreasing $\Delta T$, we found opposite trends in skewness and standard deviation for the polar and equatorial data. A smaller $\Delta T$ (i.e. a slower jet) broadens the distribution, but due to a weaker baroclinic eddy forcing the distribution becomes more narrow. It might be these opposing effects that are responsible for the observed opposing trends in the experimental distributions for mid-latitude/polar and equatorial data. Another reason might be that the baroclinic waves weaken in the vicinity of the warm outer wall, and the data from this region might capture other boundary layer related effects which impacts the statistical properties of the data.

We think the results of the study underpin the usefulness of the laboratory approach to understand specific processes of climate change in particular with a view to extreme events. However, we have only taken a first step and more sophisticated aspects like the use of extreme value theory, long term memory effects, heavy tails in the amplitude of fluctuations, power-laws, spatial correlations and teleconnections etc. have been neglected in the work presented here. Moreover, a more recent and bigger rotating tank has proven to be closer to the atmospheric case than the smaller system used here \citep{Rodda2019a}. Hence, further experiments using this bigger differentially heated rotating tank and a deeper statistical analysis are planned for the future to add more experimental data to observations and climate simulations.

\bibliographystyle{spbasic}      
\bibliography{bibliografia}   

\end{document}